%% file: main.tex
\documentclass[sigconf, nonacm,balance=false]{acmart}

\usepackage[T1]{fontenc}

\usepackage{subcaption}
\usepackage{pifont}
\usepackage{array} 
\usepackage{xurl}
\usepackage{multicol}
\usepackage{multirow}
\usepackage{makecell} 

\usepackage{float}
\usepackage{textcomp}

\usepackage{adjustbox}
\usepackage{enumitem}
\usepackage{booktabs}

\usepackage{tikz}
\usetikzlibrary{arrows.meta, positioning, shapes.multipart, fit,calc, shadows}
\usepackage{amsmath}
\usepackage{hyperref} %
\usepackage[capitalise,noabbrev]{cleveref}

\usepackage{algorithmic}
\usepackage{graphicx}
\usepackage{bmpsize}
\usepackage{xcolor}
\usepackage{lipsum}
\def\BibTeX{{\rm B\kern-.05em{\sc i\kern-.025em b}\kern-.08em
    T\kern-.1667em\lower.7ex\hbox{E}\kern-.125emX}}

\usepackage{listings}

\usepackage{xspace}
\usepackage{tikz}
\usepackage{fontawesome5}
\usetikzlibrary{arrows.meta,positioning,fit,calc}

\usepackage{tabularx}
\newcolumntype{L}{>{\raggedright\arraybackslash}X}

\AtBeginDocument{%
  \providecommand\BibTeX{{%
    Bib\TeX}}}

\setcopyright{none}
\acmConference[WWW '27]{Proceedings of the ACM Web Conference 2027}{May 10--14, 2027}{Dublin, Ireland}
\acmYear{2027}

\newcommand{\chr}{Chromium\xspace}
\newcommand{\chrs}{Chromium-S\xspace}
\newcommand{\ff}{Firefox\xspace}
\newcommand{\ffs}{Firefox-S\xspace}
\newcommand{\dom}{browser-use\xspace}
\newcommand{\gui}{SuperScraper\xspace}
\newcommand{\pp}{\,pp\xspace}
\newcommand{\ci}[2]{\,[#1,\,#2]}
\newcommand{\rqref}[1]{\textbf{RQ#1}}

\newcommand{\takeaway}[1]{\par\smallskip\noindent\fcolorbox{black!40}{black!4}{\parbox{\dimexpr\columnwidth-2\fboxsep-2\fboxrule\relax}{\textbf{Takeaway.} #1}}\par\smallskip}

\newif\ifshownotes \shownotestrue

\definecolor{changed}{HTML}{0B57A4}
\newif\ifshowchanges \showchangestrue
\DeclareRobustCommand{\chg}[1]{\ifshowchanges{\color{changed}#1}\else#1\fi}

\begin{document}

\title[Beyond the Landing Page]{Beyond the Landing Page: Website Blocking, Tracker Visibility, and Human Alignment in Automated Web Crawls}

\author{
A H M Nazmus Sakib\textsuperscript{1},
Mahsin Bin Akram\textsuperscript{1},
Raveen Wijewickrama\textsuperscript{1},
Igor Bilogrevic\textsuperscript{2},
Murtuza Jadliwala\textsuperscript{1}\\
\textsuperscript{1}The University of Texas at San Antonio \\
\textsuperscript{2}Google LLC
}
\renewcommand{\shortauthors}{Sakib et al.}

\input{abstract}

\ccsdesc[500]{Security and privacy~Web application security}
\ccsdesc[300]{Information systems~Web mining}

\keywords{web crawling, agentic crawlers, web tracking, web measurement}

\maketitle

\input{introduction}

\input{background}

\input{methods_shortened}

\input{results_shortened2_revised}

\input{discussion_revised}

\input{ethics}

\input{conclusion}

\bibliographystyle{ACM-Reference-Format}
\bibliography{references}

\appendix
\crefalias{section}{appendix}
\input{appendix_shortened}

\end{document}

%% file: abstract.tex
\begin{abstract}
Web tracking studies largely rely on automated crawlers, yet their measurements can differ from what people encounter while browsing. Conventional crawlers follow predefined navigation rules, whereas LLM-driven agents decide how to navigate and interact with websites, potentially changing the tracking they encounter. We compare agentic and non-agentic crawlers to examine how websites respond to them, what tracking they record, and how closely their observations match human browsing. Across 2,000 popular websites, we evaluate different navigation strategies, automation frameworks, and stealth configurations, alongside browsing data from 26 participants. We find that website denials increase beyond landing pages for standard browser automation, while a detection-resistant automation framework largely prevents this escalation without increasing recorded tracking. On the same URLs, agentic crawlers record tracking activity from 1.5--1.7 times as many third-party organizations, particularly on consent-managed pages. Agentic crawlers recover 82--87\% of tracker organizations observed by participants, compared with 58--68\% for non-agentic crawlers, yet roughly half of their own observations are absent from human participant visits. These findings show that avoiding website interventions, increasing tracking visibility, and achieving closer agreement with human observations do not necessarily go hand in hand, highlighting the need to select and validate crawlers according to the measurement objective.
\end{abstract}

%% file: introduction.tex
\section{Introduction}
\label{sec:intro}

Tracking is pervasive on the Web, raising concerns about users' privacy and the collection of their browsing data. Understanding the extent of Web tracking is important for assessing these risks and evaluating protections. Yet measuring what people actually encounter while browsing remains difficult.
Most large-scale studies instead rely on automated crawlers ~\cite{englehardt2016online, acar2014webneverforgets,iqbal2021fpinspector,vekaria2025sok}. A crawler, however, is not a neutral observer: its browser configuration, automation framework, and browsing behavior can affect how websites respond and what tracking activity it records~\cite{ahmad2020apophanies,jueckstock2021realistic, cassel2022omnicrawl}. Consequently, automated measurements can differ substantially from human browsing~\cite{zeber2020representativeness}.
A recent comparison with study participants found that an automated crawl missed fingerprinting on 45\% of the websites where participants encountered it, a gap that the authors attributed to authentication, bot detection, and scripts that run only after user interaction~\cite{annamalai2025beyond}.

Researchers have sought to reduce these discrepancies by simulating human interactions~\cite{goessen2021hlisa}, using realistic browser configurations~\cite{jueckstock2021realistic,cassel2022omnicrawl}, and reducing detectable automation artifacts~\cite{krumnow2022gullible}. More recently, LLM-driven browsing agents have emerged that interpret Web pages and decide how to navigate and interact with them~\cite{zhou2024webarena,he2024webvoyager, zheng2024seeact}. 
Unlike traditional crawlers that follow predefined navigation rules, these agents can make context-dependent decisions about which links to follow and which elements to interact with, potentially resembling aspects of human browsing.
These differences raise two questions: \emph{how does agentic browsing affect the tracking activity that crawlers record, and how do the resulting measurements compare with what people encounter when browsing the Web?}

Answering these questions requires distinguishing three related but easily conflated aspects of Web tracking measurement: \emph{website interventions}, the challenges and denials crawlers encounter; \emph{tracking visibility}, the third-party tracking and fingerprinting a crawler records; and \emph{human agreement}, how those observations compare with what people encounter on the same pages. A crawler can improve one of these properties without the others: it may avoid every intervention and still trigger little tracking, or it may record more tracking than any visitor encounters. 
We therefore investigate agentic crawlers across these three dimensions, comparing their observations with those of non-agentic crawlers:

\begin{itemize}[leftmargin=*,nosep]
\item \rqref{1} (Website interventions): How do the website interventions that agentic and non-agentic crawlers encounter differ, particularly beyond landing pages, and which configuration components account for the differences?
\item \rqref{2} (Tracking visibility): On matched URLs, how do the third-party tracking and browser fingerprinting that agentic and non-agentic crawlers observe differ, and what factors account for these differences?
\item \rqref{3} (Human agreement): How well do agentic and non-agentic crawlers recover the tracking and fingerprinting activity that participants observed, and what additional activity do they record?
\end{itemize}

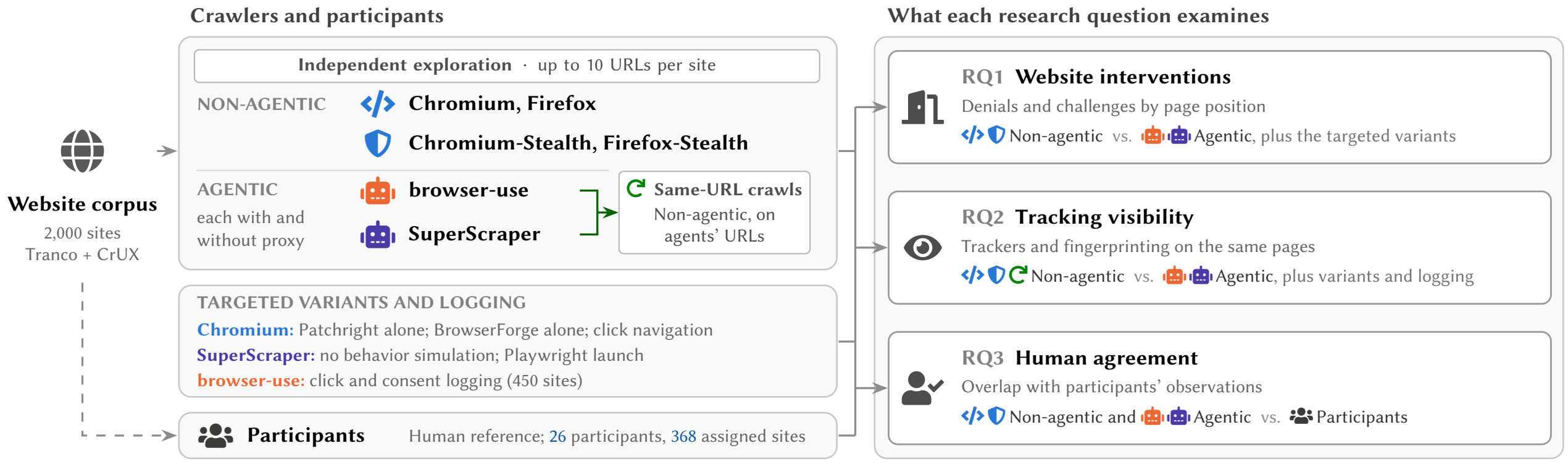
\begin{figure*}[t]
\centering
\scalebox{0.94}{%
    \input{figures/fig_overview}
}
\caption{Study design overview.}
\label{fig:overview}
\end{figure*}

To address these RQs, we study crawling on 2,000 popular websites (\Cref{fig:overview}), comparing four non-agentic and two agentic crawlers. The non-agentic crawlers use Playwright, a browser automation framework, to drive Chromium or Firefox. Their stealth variants incorporate browser fingerprint spoofing, with the Chromium variant also using Patchright, a modified version of Playwright designed to reduce automation detection. The two agentic crawlers are an agent built on the \texttt{browser-use} framework~\cite{browseruse} and \gui, a research crawler we built that perceives pages only through screenshots and acts through operating-system input.
Because each crawler chooses its own pages, a non-agentic crawler revisited every URL that the agents had visited, which lets us compare crawlers on identical pages.
Five targeted variants of the Chromium and \gui crawlers isolate the effects of specific configuration choices. An additional interaction-logging crawl records the \texttt{browser-use} agent's clicks and subsequent third-party script activity. Finally, browsing data from 26 participants on a subset of the sites provide a human reference.

Our results reveal several ways in which crawler design affects measurement outcomes. First, website denials are similar across configurations on landing pages but increase on subsequent pages for browsers driven by unmodified Playwright. Replacing Playwright with Patchright largely eliminates this escalation, whereas browser fingerprint spoofing and simulated mouse movement do not. Yet Patchright does not increase recorded tracking, demonstrating that avoiding interventions and increasing tracking visibility are distinct.
Second, on matched URLs, the agentic crawlers record trackers from 1.5--1.7 times as many organizations as non-agentic crawls. This difference is particularly pronounced on consent-managed pages. The interaction-logging crawl shows that our \texttt{browser-use} agent frequently accepted consent and that confirmed consent actions were followed by additional third-party scripts. 
Finally, the agentic crawlers recover 82--87\% of the tracker organizations observed by participants on shared URLs, compared with 58--68\% for non-agentic crawlers. However, only 48--49\% of the agents' own observations also appear in the participant visits. Thus, greater tracking visibility does not necessarily translate into closer agreement with human browsing. 

\noindent In summary, our contributions are:
\begin{itemize}[leftmargin=*,nosep]
\item \textbf{A systematic comparison of agentic and non-agentic Web crawlers.} We evaluate crawler configurations across 2,000 websites, examining how navigation strategies, automation frameworks, browsing behavior, and browser environments affect website interventions and tracking visibility.

\item \textbf{An evaluation against human browsing.} We compare agentic and non-agentic crawler observations with participants' visits to shared URLs, distinguishing the recovery of participant-observed tracking from additional activity absent from those visits.

\item \textbf{Practical guidelines and reusable artifacts for Web tracking measurement.} We derive recommendations for selecting, configuring, and validating crawlers according to different measurement objectives. To support reproducibility, we will release \gui, our crawler configurations, measurement code, analysis code, and per-page crawl data.

\end{itemize}

%% file: figures/fig_overview.tex
\begin{tikzpicture}[
  font=\sffamily\scriptsize,
  >={Stealth[length=1.8mm]},
  ttl/.style={font=\sffamily\footnotesize\bfseries, text=black!75, anchor=base west, inner sep=0pt},
  grp/.style={font=\sffamily\scriptsize\bfseries, text=black!50, anchor=base west, inner sep=0pt},
  ico/.style={anchor=center, inner sep=0pt},
  nm/.style={font=\sffamily\footnotesize\bfseries, anchor=base west, inner sep=0pt},
  dsc/.style={font=\sffamily\scriptsize, text=black!60, anchor=base west, inner sep=0pt},
  tag/.style={draw=black!30, fill=white, rounded corners=1.5pt, inner sep=2.5pt,
              font=\sffamily\scriptsize, text=black!70, align=center},
  panel/.style={draw=black!22, fill=black!2.5, rounded corners=4pt, line width=0.5pt},
  rq/.style={draw=black!40, fill=white, rounded corners=3pt, line width=0.55pt,
             minimum width=7.55cm, minimum height=1.28cm},
  flow/.style={->, line width=0.65pt, draw=black!45},
  branch/.style={->, line width=0.65pt, draw=black!45, dashed}
]
\definecolor{cSS}{HTML}{4A3AA7}
\definecolor{cDOM}{HTML}{EB6834}
\definecolor{cChr}{HTML}{2A78D6}
\definecolor{cRep}{HTML}{008300}
\definecolor{cHum}{HTML}{3A3A38}

\node[ico, font=\huge, text=black!65] at (0.95,0.55) {\faIcon{globe}};
\node[align=center, anchor=north, inner sep=0pt] at (0.95,0.05) {
  {\footnotesize\bfseries Website corpus}\\[2pt]
  {\color{black!60}2,000 sites}\\
  {\color{black!60}Tranco + CrUX}};

\node[ttl] at (2.17,2.02) {Crawlers and participants};
\path[panel] (2.05,-0.80) rectangle (9.55,1.85);

\node[tag, anchor=north west, text width=6.95cm] at (2.22,1.71) {
  \textbf{Independent exploration}\enspace$\cdot$\enspace up to 10 URLs per site};

\node[grp] at (2.25,1.02) {NON-AGENTIC};
\node[ico, font=\normalsize, text=cChr] at (4.32,1.09) {\faIcon{code}};
\node[nm] at (4.66,1.02) {Chromium, Firefox};
\node[ico, font=\normalsize, text=cChr] at (4.32,0.64) {\faIcon{shield-alt}};
\node[nm] at (4.66,0.57) {Chromium-Stealth, Firefox-Stealth};

\draw[black!15, line width=0.4pt] (2.25,0.33) -- (6.95,0.33);

\node[grp] at (2.25,0.05) {AGENTIC};
\node[dsc] at (2.25,-0.27) {each with and};
\node[dsc] at (2.25,-0.53) {without proxy};
\node[ico, font=\normalsize, text=cDOM] at (4.32,0.10) {\faIcon{robot}};
\node[nm] at (4.66,0.03) {browser-use};
\node[ico, font=\normalsize, text=cSS] at (4.32,-0.40) {\faIcon{robot}};
\node[nm] at (4.66,-0.47) {SuperScraper};

\node[tag, anchor=west] (sameurl) at (7.06,-0.15) {
  {\color{cRep}\faIcon{redo}}\enspace\textbf{Same-URL crawls}\\[1pt]
  Non-agentic, on\\agents' URLs};
\draw[line width=0.65pt, draw=cRep!75!black]
  (6.62,0.10) -- (6.82,0.10) -- (6.82,-0.40) -- (6.62,-0.40);
\draw[flow, draw=cRep!75!black] (6.82,-0.15) -- (sameurl.west);

\path[panel] (2.05,-2.25) rectangle (9.55,-0.98);
\node[grp] at (2.25,-1.24) {TARGETED VARIANTS AND LOGGING};
\node[dsc] at (2.25,-1.55) {{\color{cChr}\textbf{Chromium:}} Patchright alone; BrowserForge alone; click navigation};
\node[dsc] at (2.25,-1.84) {{\color{cSS}\textbf{SuperScraper:}} no behavior simulation; Playwright launch};
\node[dsc] at (2.25,-2.13) {{\color{cDOM}\textbf{browser-use:}} click and consent logging (450 sites)};

\path[panel] (2.05,-2.95) rectangle (9.55,-2.43);
\node[ico, font=\normalsize, text=cHum] at (2.47,-2.69) {\faIcon{users}};
\node[nm] at (2.82,-2.76) {Participants};
\node[dsc] at (4.66,-2.76) {Human reference; \chg{26} participants, \chg{368} assigned sites};

\def\iNon{{\color{cChr}\faIcon{code}\,\faIcon{shield-alt}}}
\def\iAg{{\color{cDOM}\faIcon{robot}}\,{\color{cSS}\faIcon{robot}}}
\def\iRep{{\color{cRep}\faIcon{redo}}}
\def\iHum{{\color{cHum}\faIcon{users}}}
\def\vs{{\color{black!45}\enspace vs.\enspace}}

\node[ttl] at (10.12,2.02) {What each research question examines};
\path[panel] (9.98,-2.95) rectangle (17.83,1.85);

\foreach \y/\icon/\num/\heading/\detail/\who [count=\i] in {
   1.05/door-open/RQ1/{Website interventions}/{Denials and challenges by page position}/
     {\iNon\,Non-agentic\vs\iAg\,Agentic{\color{black!55}, plus the targeted variants}},
  -0.55/eye/RQ2/{Tracking visibility}/{Trackers and fingerprinting on the same pages}/
     {\iNon\,\iRep\,Non-agentic\vs\iAg\,Agentic{\color{black!55}, plus variants and logging}},
  -2.15/user-check/RQ3/{Human agreement}/{Overlap with participants' observations}/
     {\iNon\,Non-agentic and \iAg\,Agentic\vs\iHum\,Participants}
} {
  \node[rq, anchor=west] (rq\i) at (10.13,\y) {};
  \node[ico, font=\Large, text=black!70] at (10.53,\y) {\faIcon{\icon}};
  \node[nm] at (10.96,\y+0.27) {{\color{black!50}\num}\enspace\heading};
  \node[dsc] at (10.96,\y-0.05) {\detail};
  \node[anchor=base west, inner sep=0pt, text=black!80] at (10.96,\y-0.39) {\who};
}

\draw[flow] (1.80,0.55) -- (2.03,0.55);
\draw[branch] (0.95,-0.95) -- (0.95,-2.69) -- (2.03,-2.69);

\draw[line width=0.65pt, draw=black!45] (9.76,1.05) -- (9.76,-2.69);
\foreach \y in {0.55,-1.62,-2.69}
  \draw[line width=0.65pt, draw=black!45] (9.57,\y) -- (9.76,\y);
\foreach \i in {1,...,3}
  \draw[flow] (9.76,0 |- rq\i.west) -- (rq\i.west);

\end{tikzpicture}

%% file: background.tex
\section{Related Work}
\label{sec:related}

\noindent\textbf{Validity of crawl-based measurement.} Studies of tracking and fingerprinting rely on instrumented crawls~\cite{englehardt2016online,acar2014webneverforgets,nikiforakis2013cookieless} and on heuristics or classifiers applied to recorded browser API calls~\cite{iqbal2021fpinspector,bahrami2022fpradar,laperdrix2020survey}. A growing body of work questions how well such crawls represent real browsing: automated crawls differ systematically from real users' telemetry~\cite{zeber2020representativeness}, common practices such as Selenium automation change what websites serve~\cite{cassel2022omnicrawl}, crawler design and configuration alter measurement outcomes~\cite{ahmad2020apophanies,jueckstock2021realistic}, and internal pages differ from landing pages in structure and third-party activity~\cite{urban2020beyond,aqeel2020landing}. Instrumentation choices, such as recording JavaScript inside the browser runtime rather than through page-level hooks, also shape results and their reproducibility~\cite{jueckstock2019visiblev8,demir2022reproducibility}. Closest to our work, Annamalai et al.\ found that an automated crawl of the pages that 30 participants had browsed missed 45\% of the websites on which the participants encountered fingerprinting~\cite{annamalai2025beyond}. We extend this comparison to agentic crawlers and separate the reasons why a crawler records more or less than people do.

\noindent\textbf{Bot detection as a measurement threat.} Websites detect automation through browser fingerprints~\cite{jonker2019fingerprint,vastel2020fpcrawlers}, behavioral signals~\cite{iliou2021advancedbots}, and network characteristics~\cite{jarad2026tls}, often through commercial services~\cite{azad2020webrunner}, and because anti-abuse systems fingerprint clients to classify them~\cite{bursztein2016picasso}, the fingerprinting that a crawler records can depend on whether a site considers it suspicious. Measurement crawlers have therefore adopted human-like interaction~\cite{goessen2021hlisa} and been audited for detectable artifacts~\cite{krumnow2022gullible}, and honeysites have characterized automated traffic~\cite{li2021goodbot}. Gundelach et al.\ measured blocking on the homepages of the Tranco top 10K under four Playwright configurations and found it concentrated among particular infrastructure providers~\cite{gundelach2026detecting}. We show that homepage measurements hide most of the differences between configurations, which emerge on later pages, and that reducing blocking does not by itself increase the tracking recorded.

\noindent\textbf{LLM browsing agents and consent.} Browsing agents act on DOM-derived representations, screenshots, or both~\cite{deng2023mind2web,zhou2024webarena,he2024webvoyager,zheng2024seeact,cheng2024seeclick,lu2024omniparser}, and they are evaluated on task completion rather than on what they are exposed to while browsing. Recent work studies agents from the website's perspective, distinguishing them from humans and from one another through fingerprints and interaction traces~\cite{fayolle2026nobody,wang2026fpagent,lugoloobi2026knownactions} and proposing defenses against agent-driven scraping~\cite{li2026webcloak}. Agents also make consent decisions: in controlled settings they frequently accept cookie banners on the user's behalf~\cite{saito2026invisible}, and accepting consent increases third-party cookies~\cite{rasaii2023cookieverse} and propagates consent signals to many vendors~\cite{matte2020cookie}. We measure what instrumentation records during agents' visits to live websites and quantify how their consent clicks change the scripts that load.

%% file: methods_shortened.tex
\section{Study Design}
\label{sec:design}
We evaluate eight main crawler configurations: four non-agentic and four agentic. The latter correspond to two agentic crawlers, each run with and without a residential proxy. We also conduct same-URL crawls, test targeted configuration changes, and record agent interactions to investigate differences between crawlers. Browsing data from human participants provide a reference for evaluating their observations. \Cref{fig:overview} summarizes the study design.

\subsection{Sites and Crawler Configurations}
\label{sec:configs}

\noindent\textbf{Sites.} 
We constructed a 2,000-site corpus by selecting 1,000 domains from the Tranco top 10K~\cite{lepochat2019tranco} and 1,000 from the CrUX global top list~\cite{cruxtoplists}, both obtained in July 2026.
We exclude adult, gambling, advertising, malicious, and non-user-facing domains (\Cref{app:metrics}). For smaller experiments, we selected a 450-site informative subset for measurement feasibility where \gui's crawl successfully retrieved at least one same-site HTML page and recorded instrumented JavaScript calls from scripts associated with at least one known tracker organization.
Using a fixed seed, we sampled with approximately equal quotas across strata based on list rank, bot-protection indicators such as detected Web Application Firewall (WAF) vendors and requests to known bot-management or CAPTCHA services, and observed tracker counts, redistributing shortfalls from smaller strata. This subset is used for the \gui variants, interaction logging, and participant assignments.

\noindent\textbf{Non-agentic crawlers.} 
We evaluate four non-agentic crawler configurations: \chr, \ff, Chromium-Stealth (\chrs), and Firefox-Stealth (\ffs). All four explore same-site links using a first-in, first-out queue, starting each site in a fresh, headed browser context. We use Playwright~\cite{playwright} to control \chr and \ff. For \chrs, we use Patchright~\cite{patchright}, a modified version of Playwright designed to reduce automation artifacts, together with a BrowserForge~\cite{browserforge} desktop profile and browser-property overrides. For \ffs, we use a generated profile and the same overrides, retaining Playwright after encountering compatibility issues with Firefox-specific alternatives such as Camoufox ~\cite{camoufox2026}.
We disable Firefox's built-in tracking protections in both Firefox configurations to improve comparability with Chromium.

We selected Patchright as a representative, reproducible stealth framework based on its public availability, adoption, and active maintenance~\cite{patchright2026release} at the time of data collection. We prioritized maintainability over maximizing evasion, as detection vendors can identify and target stealth frameworks' fingerprints~\cite{goessen2021hlisa,wang2026fpagent}. Additional execution details and browser preferences appear in \Cref{app:agents}.

\noindent\textbf{Agentic crawlers.} 
We evaluate two agentic crawlers: \dom, built on \texttt{browser-use} 0.13.4~\cite{browseruse}, and our screenshot-driven research crawler \gui (designed strictly as a browsing agent for tracking measurement rather than content extraction). Both use Qwen3-VL-8B~\cite{qwen3vl2025}, which we instruct to visit ten distinct pages per site, favor internal links, and dismiss obstructing overlays. The \dom agent receives DOM-derived page text and indexed interactive elements, can request screenshots, and controls Chromium through direct Chrome DevTools Protocol (CDP) commands rather than Playwright. In contrast, \gui perceives pages through screenshots annotated by an OmniParser-based pipeline~\cite{lu2024omniparser} and interacts through operating-system mouse and keyboard input with randomized timing; no browser automation framework controls its navigation (\Cref{app:agents}). We run both agents with and without Webshare residential proxies~\cite{webshare} to isolate the effect of network origin on tracking visibility and website interventions.

\subsection{Data Collection}
\label{sec:data_collection}

\noindent\textbf{Main crawls.} We run each of the eight configurations on the 2,000-site corpus, targeting up to ten distinct pages per site. Each crawler explores sites independently, so the resulting crawls may visit different URLs.

\noindent\textbf{Same-URL crawls.} 
Because crawlers independently select which links to follow, they may explore different URLs within the same website.
Therefore, to compare agentic and non-agentic crawlers on the same URLs, we adapt the URL-revisiting approach of Annamalai et al.~\cite{annamalai2025beyond}, conducting two non-agentic Chromium crawls, each revisiting the URLs recorded by one of the agents.
Each URL is loaded as a separate one-page task in a fresh browser context. Landing pages are visited first, with internal URLs revisited a median of 8--12 hours later. These crawls hold the
requested URLs fixed, but not prior navigation,
consent state, dwell time, or interactions (\Cref{sec:engagement}).

\noindent\textbf{Targeted variants.} 
To isolate crawler configuration choices, we evaluate three \chr variants on the 2,000-site corpus: replacing Playwright with Patchright, using a BrowserForge-generated profile with Playwright, and replacing queued-URL loads with visible-link clicks via GhostCursor~\cite{pyghostcursor}. On the 450-site subset, we evaluate two \gui variants: one without behavior simulation and another using Playwright to launch the browser while retaining operating-system input for navigation. 
We compare the \chr variants against their main crawl. Because \gui's main crawl was conducted substantially earlier, we use a separate unmodified \gui crawl on the same sites as the reference for its variants (\Cref{tab:overview}).

\noindent\textbf{Interaction-logging crawl.}
To examine agent interactions and subsequent third-party script loading, we ran an additional \dom crawl on the 450-site subset, using the same model and browsing task as the main crawl. We obtained usable interaction logs from 445 sites. We classify consent-related clicks as \emph{accept}, \emph{reject}, \emph{settings}, or \emph{weak accept} (e.g., ``OK''). A consent click is considered \emph{confirmed} if a consent-state change occurs in the same tab within 15\,s, indicated by a TCF (Transparency and Consent Framework) \texttt{useractioncomplete} event, Google Consent Mode update, or first write of a known consent cookie. We count newly observed third-party script domains within 10\,s of each click, comparing consent-related and ordinary clicks. Further details on logging and validation appear in \Cref{app:agents}.

\noindent\textbf{Participants.} With IRB approval and informed consent, we recruited 26 adults (18 through CloudResearch Connect~\cite{cloudresearchConnectRecruit}, 8 local university students), compensating each with \$10. 
We assigned each participant 15 sites from the 450-site subset and asked them to browse naturally, visiting at least ten distinct subpages per site. Participants used Chrome with browsing data cleared and other extensions disabled, alongside a custom-built measurement extension that collected the same logging data as the crawlers (\Cref{app:agents}).
The resulting logs contain 4,353 pages across 415 sites, including 368 assigned sites (and 6 additional unassigned sites) with recorded landing-page visits. Not all participants completed every assigned site or reached the requested ten subpages per site. We pool participants' observations as a descriptive human reference, rather than an estimate of population-level browsing behavior or individual browsing patterns.

\noindent\textbf{Collection period.}
Data collection spans July--October 2026. \Cref{tab:overview} summarizes the collection periods and coverage. Because the crawls were conducted at different times, we examine repeated collections to assess temporal variation (\Cref{tab:drift}).

\input{tables/tab_overview_rows.tex}

\subsection{Instrumentation and Data Preparation}
\label{sec:instr}

We use OpenWPM-derived JavaScript instrumentation~\cite{englehardt2016online} across all crawls and participant browsing sessions. The instrumentation is byte-identical across Chromium-based configurations and adapted for Firefox with the same hooks. It records calls to fingerprinting-relevant APIs, their arguments (except for \texttt{getContext}), and calling-script URLs, but not return values. We also record visit starts, intervention signals, bot-management headers, and eligible HTTP 2xx responses (details in \Cref{app:metrics}).
\gui and participants use a Chrome extension with debugging-interface response capture, \dom uses CDP, and the non-agentic crawlers use their respective automation frameworks.

We match pages by site and normalized requested URL, retaining the first eligible attempt (\Cref{app:metrics}). However, we noticed that response capture is missing on approximately 85\% of \gui crawled landing pages (accounting for nearly all of its 29–32\% overall missing rate in \Cref{tab:overview}), and even nominally captured pages contained little usable network data. We therefore exclude \gui's landing pages from network-derived comparisons, while retaining separately recorded intervention signals and JavaScript calls. \Cref{app:metrics} provides further details on data preparation, capture failures, and sensitivity checks.

\subsection{Metrics}
\label{sec:metrics}

\noindent\textbf{Website interventions.} We measure website interventions using page denials, hard blocks,
and challenge markers. We mark a page as \emph{denied} when we
record a first-party HTTP 403 (Forbidden) or 429 (Too Many Requests)
intervention event whose URL matches the requested page. The
\emph{denial rate} is the fraction of attempted pages classified
as denied. This metric captures recorded 403/429 denials;
blocking conveyed through other status codes or page content
is not counted as a page denial.
A \emph{hard block} is any recorded first-party 403 or 429 response
during a visit, including responses to requests other than the
requested page. A \emph{challenge marker} is a page element
associated with a known challenge or bot-management provider
(e.g., a CAPTCHA), classified as \emph{visible} when displayed
to the visitor and \emph{passive} when hidden
(e.g., invisible bot checks).
We define \emph{escalation} as the mean across sites of each site's
denial rate on pages 2--$K$ minus its landing-page rate,
considering sites with at least $K$ recorded pages
($K{=}6$ unless noted).

\noindent\textbf{Tracking visibility.} 
We count \emph{tracker organizations per page} as the distinct owners of third-party domains appearing in captured responses and classified by DuckDuckGo Tracker Radar~\cite{trackerradar} as advertising, analytics, session-replay, or social tracking. A matched pair of visits is \emph{consent-managed} if a consent-management platform (e.g., OneTrust or Didomi) appears in either visit. 
For fingerprinting, our \emph{strict} heuristic flags a page when an attributable script performs at least one of five behaviors: canvas text drawing followed by extraction, offline audio rendering followed by sample readback, WebRTC candidate gathering without media, extensive font enumeration, or querying the unmasked WebGL vendor or renderer. 
Because WebGL activity depends on browser launch settings (\Cref{sec:fingerprinting}), we also report a \emph{WebGL-free} heuristic that excludes WebGL vendor and renderer queries.
We further use a legacy fingerprinting heuristic based on Englehardt and Narayanan~\cite{englehardt2016online} as a sensitivity check (\Cref{app:metrics}).
Because fingerprinting also serves defensive purposes such as fraud and bot detection~\cite{bursztein2016picasso}, and our heuristics cannot establish a script's intent, we treat fingerprinting signals as observed behavior rather than as evidence of privacy-invasive tracking.

\noindent\textbf{Human agreement.}
On pages visited by both a crawler and participants, we compare
tracker organizations within each page and pool these observations
across pages. \emph{Recall} is the fraction of participant observations
also recorded by the crawler; \emph{precision} is the fraction of
crawler observations also recorded by participants. We additionally
report the shares of their pooled union observed only by participants,
only by the crawler, or by both.

\subsection{Comparisons and Analysis Methods}
\label{sec:stats}
We compare crawls and participant observations on \emph{matched pages}, defined as pages with the same normalized requested URL attempted by both. Network-based comparisons additionally require response capture in both visits. We report paired mean differences with equal weight per page, in percentage points (pp) for rates, and ratios of summed tracker-organization counts. We compute 95\% confidence intervals using 2,000 whole-site bootstrap resamples to account for dependence among pages from the same site. For crawler contrasts, we use site-level sign-flip tests with 10,000 random sign flips per test and apply Holm's step-down correction~\cite{holm1979simple} to adjust $p$-values within two groups of tests spanning crawler contrasts and outcome metrics (\Cref{tab:contrasts}). 
As sensitivity checks, we also examine \emph{unblocked} pairs (both visits capture the page's HTML, with neither recording a hard block or visible challenge) and \emph{landing-page} pairs (neither visit has preceding navigation within the site). Additional statistical details and assumptions appear in \Cref{app:metrics}.

%% file: tables/tab_overview_rows.tex
\begin{table}[t]
\centering
\scriptsize
\setlength{\tabcolsep}{3pt}
\caption{Collection periods (2026) and coverage.
\emph{Pages}: first attempt per requested URL;
\emph{Miss. (\%)}: pages without response capture;
\emph{Obs.\ (s)}: median time from page start to the last recorded event.}
\label{tab:overview}
\begin{tabular}{@{}llrrrrr@{}}
\toprule
\textbf{Collection} & \textbf{Dates} & \textbf{Sites} &
\textbf{Pages} & \textbf{Pg/site} & \textbf{Miss.\,(\%)} &
\textbf{Obs.\,(s)} \\
\midrule
\multicolumn{7}{@{}l}{\textbf{Main crawls}} \\
Chromium & 9/23--26 & 1,987 & 9,805 & 4.9 & 3 & 14 \\
Chromium-S & 9/27--30 & 2,000 & 9,844 & 4.9 & 4 & 9 \\
Firefox & 10/12--16 & 1,988 & 8,725 & 4.4 & 4 & 15 \\
Firefox-S & 10/12--15 & 1,981 & 8,870 & 4.5 & 3 & 16 \\
browser-use & 7/19--31 & 1,971 & 12,559 & 6.4 & 5 & 19 \\
browser-use (proxy) & 8/21--9/4 & 1,982 & 13,061 & 6.6 & 4 & 17 \\
SuperScraper & 7/15--25 & 1,988 & 5,837 & 2.9 & 29 & 26 \\
SuperScraper (proxy) & 8/21--28 & 1,996 & 6,090 & 3.1 & 32 & 24 \\

\midrule
\multicolumn{7}{@{}l}{\textbf{Same-URL crawls}} \\
Same-URL (browser-use) & 8/13--14 & 1,944 & 13,654 & 7.0 & 3 & 8 \\
Same-URL (SuperScraper) & 8/6 & 1,990 & 6,859 & 3.4 & 4 & 6 \\

\midrule
\multicolumn{7}{@{}l}{\textbf{Targeted variants and reference}} \\
Chromium $\to$ Patchright & 10/5--7 & 1,996 & 12,477 & 6.3 & 3 & 6 \\
Chromium + BrowserForge & 10/5--10 & 1,999 & 9,704 & 4.9 & 2 & 13 \\
Chromium + click navigation & 10/2--5 & 1,997 & 5,497 & 2.8 & 4 & 19 \\
SuperScraper (reference) & 10/18--19 & 450 & 1,804 & 4.0 & 25 & 24 \\
SuperScraper $-$ behavior sim. & 10/5--6 & 450 & 1,913 & 4.3 & 24 & 13 \\
SuperScraper + Playwright launch & 10/5--6 & 450 & 1,918 & 4.3 & 1 & 24 \\

\midrule
\multicolumn{7}{@{}l}{\textbf{Participant sessions}} \\
Participants & 9/11--10/18 & 415 & 4,353 & 10.5 & 8 & 7 \\

\bottomrule
\end{tabular}
\end{table}

%% file: results_shortened2_revised.tex
\section{Results}
\label{sec:results}

We present findings for each research question. \Cref{tab:contrasts} in \Cref{app:extra} reports crawler contrasts and Holm-adjusted tests.

\subsection{Website Interventions}
\label{sec:access}

\begin{figure}[htb]
\centering
\includegraphics[width=0.9\columnwidth]{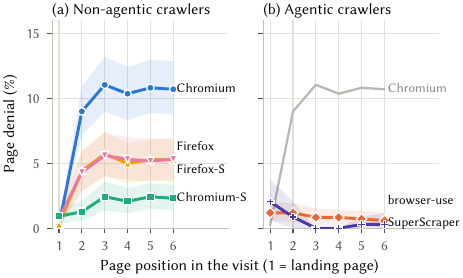}
\caption{
Page denial by position in the visit, on sites with at least six recorded pages (bands: 95\% CIs).
}
\label{fig:access}
\end{figure}

\noindent\textbf{Page denials.} On the 1,902 landing pages attempted by all eight configurations, denial rates were similar for non-agentic (6.0--8.1\%) and agentic crawlers (5.3--8.1\%); 86\% of these sites recorded no denial in any configuration (\Cref{tab:accessmerged}). Differences emerged on subsequent pages (\Cref{fig:access}). Among sites with at least six pages per crawler, \chr's denial rate increased from 0.2\% on landing pages to 10.4\% on pages 2--6, an escalation of +10.1\pp\ci{8.3}{12.1}. Escalation was smaller for \ff (+4.9\pp), \ffs (+4.4\pp), and \chrs (+1.2\pp), and absent for \dom ($-$0.4\pp) and \gui ($-$1.7\pp). Five- and eight-page analyses and hard-block measurements showed similar patterns (\Cref{tab:accessmerged}). Thus, similar denial rates on landing pages do not imply similar access during continued browsing.

\noindent\textbf{Targeted variants.} Configuration changes help explain this pattern (\Cref{tab:ablation}). Although \chr and \chrs use the same link-selection strategy, \chrs encountered fewer denials on 5,122 matched internal pages (2.4\% versus 8.5\%), despite shorter median intervals between requests (9.9 versus 14.2\,s). Replacing Playwright with Patchright alone reduced \chr's escalation from +10.1 to +1.8\pp and matched internal-page denial by 6.4\pp. BrowserForge alone (+9.3\pp) and click navigation with simulated mouse movement (+9.8\pp) retained escalation similar to \chr; \ffs likewise retained Playwright and showed escalation similar to \ff.

The two agents had similar denial rates on matched pages (difference: 0.0\pp\ci{$-$0.6}{0.5}). Launching \gui's browser through Playwright while retaining operating-system navigation did not create escalation ($-$0.4\pp), although it raised matched internal-page denial by 1.3\pp\ci{0.0}{2.9}. Together, these variants suggest that navigation through the automation framework contributes more to website detection than how the browser is launched or whether human-like interaction is simulated.

\begin{table}[htb]
\centering
\footnotesize
\setlength{\tabcolsep}{2.6pt}
\caption{
Targeted configuration variants and their effects
on website interventions, tracking visibility, and
instrumentation. SuperScraper variants use a separate
unmodified run on the same 450 sites as their reference.
}
\label{tab:ablation}
\begin{tabular}{@{}lrrrrr@{}}
\toprule
& \multicolumn{2}{c}{\textbf{Interventions (pp)}} & \textbf{Visibility} & \multicolumn{2}{c}{\textbf{Instrument (\%)}} \\
\cmidrule(lr){2-3}\cmidrule(lr){4-4}\cmidrule(lr){5-6}
Configuration & Escalation & $\Delta$\,denial & Trackers & Uncertain & WebGL \\
\midrule
\input{tables/tab_ablation_rows.tex} 
\bottomrule
\end{tabular}
\end{table}

\noindent\textbf{Cloudflare-served pages and passive challenges.} Denials were particularly concentrated on Cloudflare-served internal pages: \chr was denied on 21.6\% of these pages, versus 4.3\% elsewhere. Among Cloudflare-served pages, denial was lower for \chrs (1.7\%) and the agents (0.5--1.0\%) than for \ff and \ffs (14.6--15.1\%). Yet passive challenge markers appeared on 17--24\% of agents' matched pages versus 2--11\% of non-agentic pages, despite the agents' lower denial rates. Participants also encountered more passive markers (9--12\%) than \chr (6.1\%).

\takeaway{\emph{How browsing is automated matters more for avoiding website interventions than whether an agent directs navigation. Passive challenge markers are not reliable indicators for actual access denial.}}

\subsection{Tracking Visibility}
\label{sec:visibility}
We compare tracker organizations identified in captured third-party responses and scripts exhibiting fingerprinting behavior.

\subsubsection{Third-Party Tracking}
\label{sec:engagement}

\noindent\textbf{Tracker counts.} On 11,435 matched pages, \dom recorded 6.00 tracker organizations per page versus 4.01 for its non-agentic same-URL crawl (ratio 1.50\ci{1.42}{1.58}). On 3,759 matched internal pages, \gui recorded 7.24 versus 4.29 (ratio 1.69\ci{1.57}{1.81}); its landing pages are excluded from network comparisons (\Cref{sec:instr}). Comparisons of the main crawls on shared URLs; where both crawlers maintain stateful multi-page sessions rather than loading internal URLs in fresh contexts; also favored the agents (ratios 1.7–2.3 across all 16 pairs of agentic and non-agentic configurations). The two agents, despite different interfaces and collection mechanisms, recorded nearly identical counts on their shared pages (ratio 0.99\ci{0.95}{1.03}). On unblocked pages, the agents recorded more organizations across advertising, analytics, session replay, and social tracking; the difference also remained when counting only organizations whose scripts made instrumented JavaScript calls.

\begin{figure}[htb]
\centering
\includegraphics[width=0.99\columnwidth]{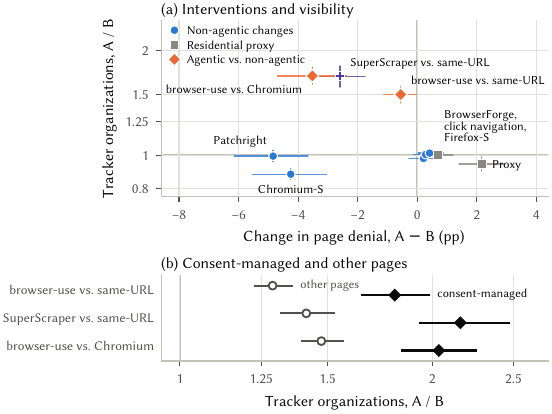}
\caption{
(a) Each paired contrast (A vs.\ B on matched pages): change in page denial against the tracker-organization ratio. (b) Agentic contrasts on pages with and without a consent-management platform. Bars: 95\% CIs.
}
\label{fig:visibility}
\end{figure}

\noindent\textbf{Beyond denials.} On matched pages where neither crawler was blocked, the agents still recorded 1.47 (\dom) and 1.62 (\gui) times as many tracker organizations as their same-URL crawls. Conversely, switching \chr to Patchright reduced matched internal-page denial by 6.4\pp without increasing tracker counts (ratio 0.99). Avoiding denials therefore does not, by itself, explain the agents' additional tracking.

\noindent\textbf{Consent-managed pages.} On matched pages where either crawler detected a consent-management platform, the agents recorded 1.80--2.16 times as many tracker organizations as their same-URL crawls, versus 1.29--1.42 times on other pages (\Cref{fig:visibility}b). Consent-managed pages comprised 28\% of matched pages but accounted for 62--65\% of the additional tracker organizations.

\noindent\textbf{Consent interactions.} To examine what happens after consent actions, we instrumented a separate \dom crawl on 445 sites. It made 273 consent-related clicks across 140 sites: 201 explicit accepts, 47 weak accepts, 23 settings openings, one rejection, and one unclassified action. 
On the 69 sites where confirmed consent clicks could be compared with ordinary page-loading clicks, the former were followed by a mean of 9.2 new third-party script domains within 10\,s, versus 1.0 (site-level means; Wilcoxon signed-rank $p<0.001$). The largest increases followed a reload or navigation after consent (\Cref{tab:logger}), and newly observed domains were predominantly advertising-related (\Cref{tab:functions}).

\begin{table}[htb]
\centering
\footnotesize
\setlength{\tabcolsep}{2.6pt}
\caption{Interaction-logging crawl (\dom, 445 sites): new third-party script domains within 10\,s of a click, grouped by the page's subsequent behavior. Clicks followed by another click within 10\,s are excluded. \emph{Any}: clicks followed by at least one new domain (\%).}
\label{tab:logger}
\begin{tabular}{@{}llrrrr@{}}
\toprule
Click & Page afterwards & Clicks & Sites & Mean & Any (\%) \\
\midrule
Confirmed consent & reloaded & 32 & 31 & 12.1 & 94 \\
 & navigated & 22 & 21 & 11.1 & 82 \\
 & unchanged & 31 & 25 & 1.7 & 35 \\
\midrule
Ordinary & navigated & 5,088 & 399 & 0.5 & 19 \\
 & reloaded & 803 & 206 & 0.4 & 13 \\
 & unchanged & 4,470 & 375 & 0.1 & 2 \\
\bottomrule
\end{tabular}
\end{table}

\noindent\textbf{Visit duration and interaction.} On matched pages, the agents had longer observation times than their same-URL crawls (median 19.6 and 23.7\,s versus 7--8\,s) and made three to eight times as many interaction-related calls. However, a \chr click-navigation variant also had longer observation times (20.9 versus 13.4\,s) and made about twice as many interaction-related calls, without recording more tracker organizations (ratio 1.00\ci{0.99}{1.02}).

\takeaway{\emph{Agents' greater tracking visibility is not simply a consequence of avoiding denials or spending more time on pages. The tracking recorded depends on which interactions occur, with consent-related actions providing one concrete explanation for the additional activity.}}

\subsubsection{Browser Fingerprinting}
\label{sec:fingerprinting}

\begin{figure*}[t]
\centering
\includegraphics[width=0.9\textwidth]{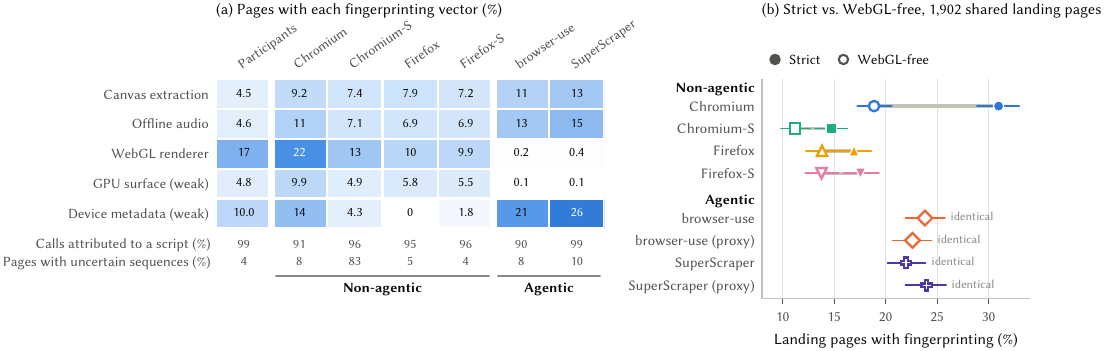}
\caption{(a) Selected fingerprinting vectors, weak indicators, and instrumentation diagnostics, computed on each configuration's own pages. (b) Strict (filled) and WebGL-free (hollow) fingerprinting prevalence on 1,902 shared landing pages; 95\% CIs.}
\label{fig:instr}
\end{figure*}

\noindent\textbf{Differences across vectors.} Unlike tracker counts, fingerprinting did not consistently increase with agentic crawling. On matched pages, offline audio fingerprinting was more frequent among agents (+2.5--4.4\pp), but WebGL renderer queries occurred on fewer than 0.5\% of agents' pages versus 10--22\% for non-agentic crawlers (\Cref{fig:instr}a). Excluding WebGL, the agents recorded only 1.0--3.2\pp more fingerprinting than \chr and their same-URL crawls, despite recording 1.5--2.2 times as many tracker organizations.

\noindent\textbf{Sources of the differences.} The agents' same-URL crawls recorded WebGL renderer queries on 16--17\% of those URLs, indicating that URL selection alone does not explain the agents' near-absence of WebGL queries. In two same-day \gui variants on 875 matched pages, WebGL queries occurred on 1.3\% with normal browser launch and no behavior simulation, versus 44.8\% with Playwright launch, while WebGL-free fingerprinting remained near 21\%. The same Google advertising hosts ran scripts in both variants but queried the renderer only under Playwright (\Cref{app:metrics}) -- consistent with defensive scripts~\cite{bursztein2016picasso,jonker2019fingerprint} probing WebGL only when browser launch flags or CDP attachment expose automation signals ~\cite{patchright,krumnow2022gullible}. Browser launch is therefore implicated, although the variants also differed in behavior simulation.

Instrumentation further complicates the comparison. The agents recorded 9.6--9.8\pp more WebGL-free fingerprinting than \chrs, versus approximately 3\pp more than \chr (\Cref{tab:contrasts}). Yet 83\% of \chrs's pages contained uncertain JavaScript call sequences, versus 8\% for \chr. Neither Patchright nor BrowserForge alone reproduced this problem (\Cref{tab:ablation}), suggesting interference from their combination when BrowserForge's prototype-spoofing wrappers and OpenWPM's hooks run together under Patchright's modified execution context, stripping call arguments.

\noindent\textbf{Consequences for comparisons.} Under the strict heuristic, \chr recorded fingerprinting on 7--8\pp more matched pages than either agent. Excluding WebGL changed the direction: both agents recorded approximately 3\pp more than \chr, a reversal also visible on the 1,902 shared landing pages (\Cref{fig:instr}b). 

\takeaway{\emph{A crawler can appear to encounter less fingerprinting because of differences in browser execution or instrumentation, rather than because the websites it visits perform less fingerprinting.}
}

\subsection{Comparison with Human Browsing}
\label{sec:human}

\noindent\textbf{Website interventions.} On participant-shared internal pages, participants were denied on at most 0.4\% of pages and the agents on none. The three configurations using standard Playwright (\chr, \ff, and \ffs) were denied on 3.9--7.2\% of these pages, compared with 1.7\% for \chrs. On matched landing pages, denial rates were 3.0\% for participants, 3.5\% for agents, and 4.3--7.8\% for non-agentic crawlers. Thus, the later-page escalation observed with standard Playwright was absent from participant visits.

\begin{figure}[htb]
\centering
\includegraphics[width=0.85\columnwidth]{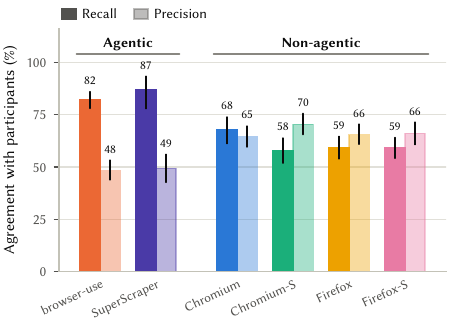}
\caption{Tracker-organization agreement with participants.
Pooled recall (solid) and precision (light) on shared
pages with network capture in both visits; 95\% CIs.}
\label{fig:human}
\end{figure}

\noindent\textbf{Tracker agreement.} On pages shared with participants, the agents recovered 82\% (\dom) and 87\% (\gui) of participant-observed tracker organizations, versus 58--68\% for non-agentic configurations (\Cref{fig:human}). However, only 48--49\% of the agents' observations were also seen in participant visits, compared with 65--70\% for non-agentic crawlers. The higher recall therefore came with lower precision, yielding similar pooled F1 scores across all six crawlers (0.61--0.66). Because \gui's landing pages lack network capture (\Cref{sec:instr}), its comparison uses only internal pages. Restricting all six crawlers to internal pages preserves the pattern (recall 85--87\% versus 57--68\%; precision 49--50\% versus 74--75\%).

\noindent\textbf{Consent-related differences.} On consent-managed pages, the agents recorded 1.83--2.00 times as many tracker organizations as participants, whereas non-agentic crawlers generally recorded fewer (0.61--0.96 times). The agents' excess persisted on other pages (1.61--1.62 times). These observations are consistent with the consent-related activity in \Cref{sec:engagement}, but participant consent decisions were not logged, and consent state may persist across pages.

\noindent\textbf{Visit duration.} Participants' median observations lasted 6.8\,s versus 26.6\,s for \dom and 5.0 versus 23.6\,s for \gui on shared pages. Truncating each agent's observations at the corresponding participant's observation duration reduced the agent-to-participant tracker ratio from 1.71\ci{1.54}{1.90} to 0.95\ci{0.84}{1.06} for \dom, and from 1.76\ci{1.49}{2.08} to 1.28\ci{1.08}{1.50} for \gui. At these durations, additional trackers were no longer evident for \dom but persisted for \gui.

\noindent\textbf{Fingerprinting agreement.} On pages shared with participants, agents recorded canvas extraction and offline audio on 14--17\% of pages, versus 5--7\% for participants and 6--10\% for non-agentic crawlers. Conversely, WebGL renderer queries appeared on at most 0.5\% of agent pages versus 17--20\% for participants. The agents' aggregate strict fingerprinting prevalence (21\%) resembled participants' (20--23\%), despite these vector-level differences. Only \chrs and the two Firefox configurations recorded WebGL-free prevalence close to participants' (8--9\% for both), although \chrs's instrumentation limitations complicate that comparison (\Cref{sec:fingerprinting}).

\takeaway{\emph{The agents recovered more participant-observed
trackers but also recorded many additional ones, while
similar aggregate fingerprinting rates concealed differences
across vectors. Thus, greater tracking visibility or similar
aggregate prevalence does not necessarily imply closer
agreement with human browsing.}
}

%% file: tables/tab_ablation_rows.tex
Chromium (reference) & +10.1 & ref. & ref. & 8.3 & 22.4 \\
\quad $\to$ Patchright & +1.8 & $-$6.4 & 0.99 & 3.7 & 14.3 \\
\quad + BrowserForge & +9.3 & +0.3 & 0.98 & 9.3 & 24.2 \\
\quad + click navigation & +9.8 & +0.5 & 1.00 & 9.1 & 18.5 \\
SuperScraper (reference) & $-$5.0 & ref. & ref. & 9.5 & 1.1 \\
\quad $-$ behavior simulation & $-$0.8 & +0.0 & 0.97 & 10.0 & 0.9 \\
\quad + Playwright launch & $-$0.4 & +1.3 & 0.93 & 10.3 & 45.4 \\

%% file: discussion_revised.tex
\section{Discussion}
\label{sec:discussion}

\noindent\textbf{Different outcomes require different controls.}
Our findings show why crawler quality cannot be reduced to a single measure of human likeness or detection avoidance. Automation frameworks affect whether a crawler can continue through a site; browsing duration and interactions influence the tracking it records; and browser execution and instrumentation affect which fingerprinting behaviors can be observed. These effects need not move together. Patchright reduced denials without increasing tracker counts, whereas the agents recovered more participant-observed trackers but also recorded substantially more activity absent from participant visits. A crawler that performs well on one measurement objective therefore need not be suitable for another.

\noindent\textbf{Choosing and validating crawlers.}
Crawler choice should follow the question being measured (\cref{tab:factors}). For studies of tracking during ordinary browsing, the visit protocol should reflect the target users' navigation, observation duration, and consent choices. Non-agentic crawlers may make those choices easier to standardize, but are not inherently representative of human visits. For studies seeking tracking activity that interaction can activate, agents can provide broader exploratory coverage, provided that their actions are logged and their observations are not presented as typical user exposure.

\begin{table}[htb]
\centering
\caption{Crawler selection and configuration parameters by measurement objective.}
\label{tab:factors}
\scriptsize
\setlength{\tabcolsep}{3pt}
\begin{tabular}{@{}>{\raggedright\arraybackslash}p{0.25\columnwidth}p{\dimexpr0.75\columnwidth-2\tabcolsep\relax}@{}}
\toprule
\textbf{Objective} & \textbf{Recommended Crawler \& Parameters} \\
\midrule
\textbf{Interaction-triggered Tracking} &
\textbf{Agentic (direct protocol control).} Autonomous exploration with overlay dismissal; log clicks across all frames and monitor consent-state APIs; capture network via browser protocol rather than extensions. \\
\addlinespace[1.5pt]
\textbf{Typical Human Tracking Exposure} &
\textbf{Non-agentic (stealth driver) or Constrained Agent.} Bound page dwell time to human medians; enforce a fixed consent policy matching target users rather than auto-accepting banners. \\
\addlinespace[1.5pt]
\textbf{Multi-page Access (low blocking)} &
\textbf{Non-agentic (stealth driver) or Agentic (protocol/OS-input).} Avoid unpatched automation drivers; crawl multiple internal pages per site; omit property spoofing and simulated mouse movement. \\
\addlinespace[1.5pt]
\textbf{Browser Fingerprinting} &
\textbf{Native-launch Browser + Unmodified Profile.} Launch browser natively rather than via automation frameworks; avoid combining driver patches with fingerprint-spoofing overrides. \\
\bottomrule
\multicolumn{2}{@{}p{\columnwidth}@{}}{\scriptsize \textbf{Essentials:} Match on normalized requested URLs; verify first-page capture completeness (notably with browser extensions); interleave runs or repeat reference crawls to bound temporal drift.}
\end{tabular}
\end{table}
\noindent Fewer denials do not guarantee complete measurements: in \chrs, arguments needed for fingerprinting classification were often missing. Likewise, identical JavaScript hooks do not guarantee comparable results if browser launch settings change which API calls scripts make. Fingerprinting studies should therefore report vector-level measurements and test instrumentation in every browser configuration, rather than relying only on aggregate prevalence. Repeated reference crawls can help assess temporal variation, but do not remove every difference between collections conducted at different times.

\noindent\textbf{Implications for the agentic Web.}
These findings also matter when agents browse on users' behalf rather than as measurement instruments. In our interaction-logging crawl, the \dom agent made 201 explicit consent-accept clicks and only one explicit rejection while following an instruction to dismiss obstructing overlays. Confirmed consent actions were followed by additional third-party script loading, and our agents recorded more tracker organizations than participants on shared pages. These observations do not establish that consent explains all of the difference or that other agents would act similarly. They do show how a seemingly minor interaction policy can affect the tracking recorded during delegated browsing. Agent frameworks should therefore make consent handling an explicit, user-controllable choice, rather than an incidental consequence of clearing obstacles to navigation. For websites, the low later-page denial rates of our agents do not establish that agentic navigation is inherently harder to detect: Patchright produced a similar reduction without using an agent, and passive challenge markers did not track denial rates consistently.

\noindent\textbf{Limitations.} We study one open model with one framework per agent design, and our crawls without a proxy ran from a single US vantage point, where consent banners are less common than in the EU, so consent effects may be larger there. The intervention results reflect stealth and detection tools at one point in time. \gui reached fewer pages per site than the other crawlers and lacks network data on most landing pages, and only the browser-use agent's consent actions were logged. The participant reference is small, pooled, and descriptive, with short visits and unrecorded consent decisions. Page denial is a lower bound on blocking, our heuristics cannot separate tracking from defensive fingerprinting~\cite{bursztein2016picasso}, and the drift over the three-month collection is bounded in \cref{sec:data_collection}.

%% file: ethics.tex
\section{Ethics}
\label{sec:ethics}

Our human-browsing study was reviewed and approved by our institution's IRB. All 26 participants provided informed consent and received \$10 for their participation. Participants used a browser extension to record Web activity for the study. The collected data included JavaScript API calls, visited URLs, and eligible HTTP responses. Participants used a separate browser profile from their personal
profile and received explicit instructions describing the data
collected. The extension was configured to initiate recording on
domains in the 2,000-site study corpus, including eligible third-party
resources loaded during those visits. We did not solicit demographic information. We will release participant observations only in aggregate, not individual browsing records.
Our automated crawlers accessed publicly available pages without credentials, visiting at most 10 pages per site per configuration sequentially within a single browser session (yielding a low per-site request rate with median inter-page intervals of 9.9–26.6 s; \Cref{sec:access,sec:visibility}) and never running concurrent sessions against the same domain. Following established Web tracking measurement methodology ~\cite{acar2014webneverforgets,englehardt2016online,jueckstock2021realistic}, our crawlers did not enforce \texttt{robots.txt} exclusions, as our objective is to measure the tracking and interventions encountered during ordinary interactive browsing rather than to index content. Agent interactions with website controls, including consent banners and a single-attempt challenge handler, occurred in isolated, stateless sessions.

%% file: conclusion.tex
\section{Conclusion}
\label{sec:conclusion}
Our findings show that agentic crawling does not uniformly improve Web-tracking measurements. Agents recorded more third-party tracking and recovered more participant-observed trackers, but also observed activity absent from human visits. Their lower denial rates could largely be reproduced with a patched automation framework, while browser configurations and instrumentation complicated fingerprinting comparisons. These findings demonstrate that website interventions, tracking visibility, and agreement with human browsing are distinct measurement objectives. We provide practical guidance for selecting and validating crawlers according to these objectives rather than assuming that any one crawler type is inherently better.

%% file: appendix_shortened.tex
\section{Crawler and Study Details}
\label{app:agents}

\begin{table*}[htb]
\centering
\footnotesize
\setlength{\tabcolsep}{3.4pt}
\caption{
All planned contrasts on matched pages (A$\,-\,$B; pp unless noted; 95\% intervals); values compare within rows only. \emph{Trackers}: organizations per page where both visits captured responses (\gui: internal pages). Bold: Holm-adjusted $p<0.05$ within its family (planned: hard block, trackers, strict; added: page denial, WebGL-free). Strict fingerprinting reflects the effects in \cref{sec:fingerprinting}.
}
\label{tab:contrasts}
\begin{tabular}{@{}lrllrlll@{}}
\toprule
& & \multicolumn{2}{c}{Interventions (pp)} & & Visibility & \multicolumn{2}{c}{Fingerprinting (pp)} \\
\cmidrule(lr){3-4}\cmidrule(lr){6-6}\cmidrule(lr){7-8}
Contrast (A $-$ B) & Pages & Page denied & Hard block & Pages & Trackers/page & Strict & WebGL-free \\
\midrule
\input{tables/tab_contrasts_rows.tex} 
\bottomrule
\end{tabular}
\end{table*}

\noindent\textbf{Non-agentic crawlers.} The queue described in \cref{sec:configs} starts at the site's root URL. For each URL, the crawler waits for document load or navigation failure, checks challenges, scrolls to the bottom, waits briefly, captures a screenshot, and queues unseen fragment-free HTTP(S) links on allowed hosts, including hosts reached through redirects. It stops after ten URLs, an empty queue, a failure limit, or the per-site time budget; a watchdog restarts unresponsive drivers. The stealth overrides cover \texttt{navigator.webdriver}, platform, hardware concurrency, device memory, notification permissions, and WebGL vendor/renderer strings. Firefox preferences disable Enhanced Tracking Protection and its social-tracking, cryptomining, fingerprinting, and email-tracking lists, bounce-tracking protection, query-parameter stripping, first-party isolation, network-state partitioning, and fingerprinting resistance. They allow all cookies, full referrers, DNS prefetching, and speculative connections; retain storage, service-worker, Beacon, and WebRTC APIs; and send neither Global Privacy Control nor Do Not Track. No non-agentic run used a proxy.

\noindent\textbf{Browser-use agent.} \dom uses headed Chromium. Its task specifies ten distinct subpages and, beyond the shared instructions (\cref{sec:configs}), avoiding repeated actions; a progress tool lists visited URLs. The driver caps steps and stops on stalled progress, errors, or the time limit.

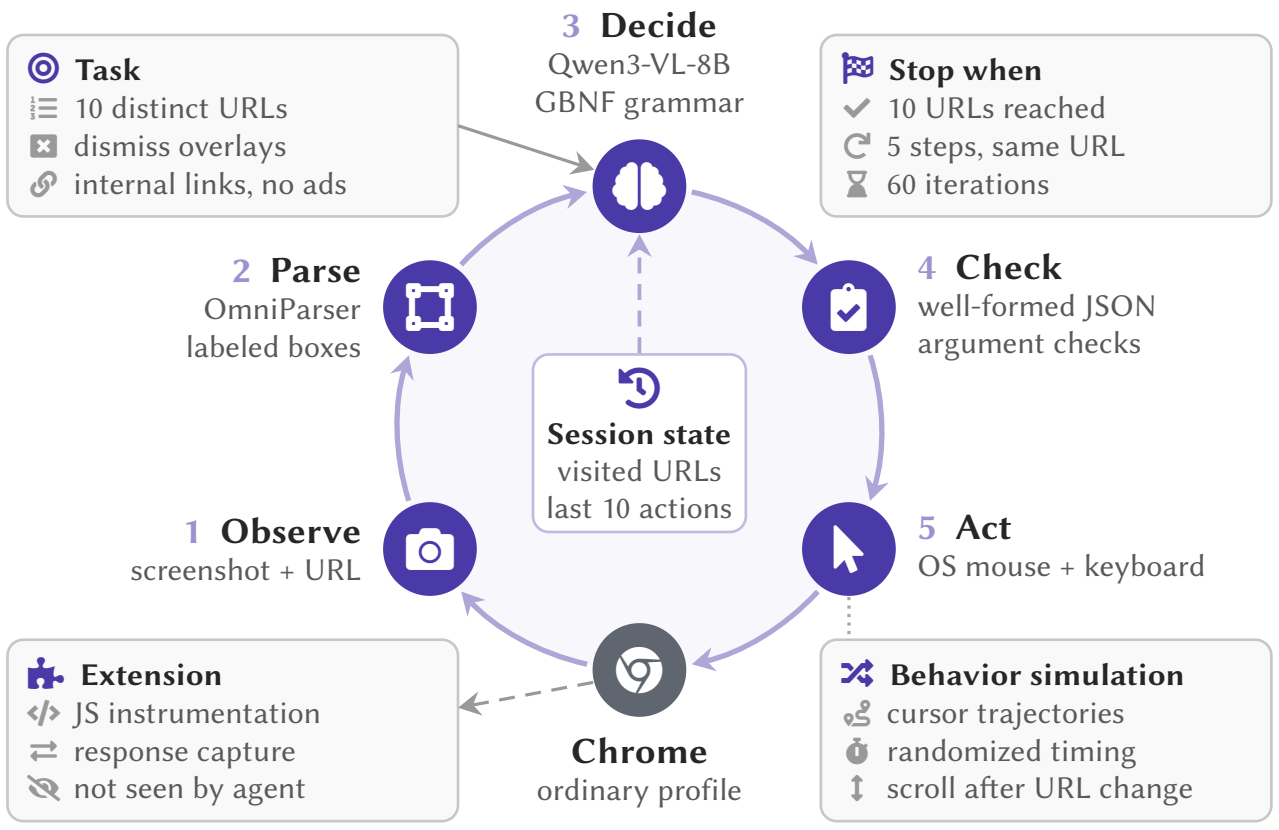
\begin{figure}[h]
\centering
\input{figures/fig_superscraper}
\caption{
\gui's perception--decision--action loop (steps 1--5, as labeled). 
}
\label{fig:superscraper}
\end{figure}

\noindent\textbf{\gui.} \gui runs a closed perception--decision--action loop in an ordinary Chrome profile (\cref{fig:superscraper}). In each iteration, it reads the URL from the address bar and captures a screenshot, which an OmniParser-based pipeline~\cite{lu2024omniparser} annotates with labeled bounding boxes through OCR, icon detection, and captioning. Qwen3-VL-8B~\cite{qwen3vl2025}, at temperature 0, receives the annotated screenshot, the element descriptions, the goal, the visited URLs, and the last ten actions, and it outputs one action in a format constrained by a GBNF grammar: a short rationale, a tool (click, double-click, move, type, enter, drag, scroll, wait, or CAPTCHA), and its arguments. Actions are executed with operating-system input, using HumanCursor~\cite{humancursor} trajectories for pointer movement and drags and PyAutoGUI~\cite{pyautogui} for keys, wheel events, and some clicks, with randomized timing: the mouse button is held for 0.2--0.5\,s, inter-key intervals last 0.05--0.2\,s, waits last 1--3\,s, and a scroll consists of 6--14 wheel events. After a URL change, with probability 0.5, the agent also moves to a random content position, scrolls down one to three times, and returns to the top. The CAPTCHA action makes one model call specialized for the challenge type and is attempted once per challenge. A site visit ends after ten distinct URLs, five consecutive steps without a URL change, or 60 iterations, and between sites the agent clears browsing data and restarts Chrome. Because screenshot-based bounding-box grounding occasionally clicks non-navigational elements or spends multiple steps scrolling and dismissing overlays without changing the URL, SuperScraper hits the five-step stall limit more often than DOM-indexed agents, resulting in fewer recorded pages per site (2.9–3.1 in \Cref{tab:overview}). The extension (\cref{sec:instr}) injects at document start in every frame; the agent never sees its output.

\noindent\textbf{Cost of agentic crawling.} Both agents ran the model locally on one \texttt{NVIDIA RTX 4000 Ada Generation} GPU with 20 GB VRAM, and every step of a visit waits for a model call. In the main runs, the recorded part of a site visit lasted a median of 3.6 minutes for \dom and 2.3 minutes for \gui, versus 1.1--2.3 minutes for the non-agentic configurations, and in the interaction-logging crawl, the median agent step took 12.8\,s and a site required a median of 39 steps. Estimated token usage was approximately 5,000 tokens per step for \gui and 12,000 for \dom. While \gui operated with a 16k-token context window, \dom required at least 22k tokens. Actual usage varies with website content and complexity.

\noindent\textbf{Interaction-logging crawl.} The interaction-logging crawl (\cref{sec:data_collection}) retained the main run's agent version and used fresh profiles and a 600\,s site limit, ending 68\% of visits before task completion. Logging reuses the agent's CDP connection, and CDP auto-attachment extends it to cross-origin iframes, which often host consent banners. Capture-phase listeners in every page and frame record clicks and form submissions together with their target elements, and 99.6\% of the recorded page clicks matched an action in the agent's step history. A click is classified as consent-related when its target lies inside a known consent-management container or frame or carries consent-related text in one of several languages. Script requests are recorded with their CDP initiator stacks. Windows that overlap the agent's next click are excluded (5.7\% of clicks at 10\,s), and the within-site comparison tests per-site means of confirmed consent clicks against those of ordinary clicks with a Wilcoxon signed-rank test; the result holds against all ordinary clicks (8.96 versus 0.42 new domains, 71 sites) and against those without a page load (8.44 versus 0.11, 63 sites), and, on the 16--21 sites whose 20\,s windows did not overlap the next click, also for 20\,s windows (all $p\le0.003$). After \texttt{doubleclick.net}, the domains most often first seen after confirmed consent were \texttt{adtrafficquality.google} (45\% of sites, versus 8\% after ordinary clicks), \texttt{googlesyndication.com} (35\% versus 7\%), \texttt{facebook.net} (34\% versus 10\%), \texttt{pubmatic.com} (31\% versus 4\%), \texttt{criteo.net} (25\% versus 4\%), \texttt{adnxs.com} (23\% versus 4\%), and the identity services \texttt{id5-sync.com} (18\% versus 4\%) and \texttt{crwdcntrl.net} (14\% versus 2\%). Fingerprinting after a click was classified per script with the strict heuristic on the calls in the 10\,s window, so a script that draws before the click and extracts after it is missed in both click groups. Without manual labels, we assessed the consent classifier against two independent signals: 68\% of the clicks it labeled as consent matched the agent's stated goal, which mentioned cookies or consent, 83\% of the accept clicks did so, and the share of likely consent clicks that it missed implies a recall of about 79\%. Confirmed clicks are a conservative subset, since unconfirmed consent clicks that were followed by a reload still brought a mean of 4.0 new domains.

\noindent\textbf{Participants.} Participants were asked to browse naturally, visit at least ten distinct subpages per assigned site, and spend a reasonable time on each. They had one week, could take breaks, and could not use VPNs, proxies, or incognito mode, or clear browsing data during the study. Logs also include sites reached beyond their assignments (\cref{sec:data_collection}).

\section{Data Preparation and Metrics}
\label{app:metrics}

\noindent\textbf{Records and matching.} Captured 2xx bodies exclude media, fonts, octet streams, PDFs, archives, and responses with declared length above 256\,KiB. Intervention logs include 403, 404, 429, and 5xx statuses, connection failures, and challenge/provider DOM markers with visibility flags. We recompute first-party status from URLs, assign visits to site tasks using session order and root-URL visits, discard other-site documents (mostly frames), and retain the first remaining attempt. Normalization lowercases scheme/host, removes fragments/default ports, merges HTTP/HTTPS, leading \texttt{www.}, and trailing slashes, but retains query strings. Matching uses requested URLs, not redirect destinations.

Corpus filtering (\cref{sec:configs}) uses StevenBlack hosts lists~\cite{stevenblackhosts}, Cloudflare family-filtering DNS~\cite{cloudflareIntroducing1111}, and domain-name heuristics. The 25 public-suffix entries (e.g., \texttt{github.io}, \texttt{blogspot.com}, \texttt{myshopify.com}) cover their subdomains: 465/74,791 pages (0.6\%). Excluding them changes headline tracker ratios by less than 0.01 and escalation by less than 0.1\pp.

\noindent\textbf{\gui's response capture.} Response capture is missing on 1,684 of \gui's 1,988 landing pages (84.7\%), versus 0.9\% of its internal pages, and on 95\% of the landing pages of its proxy run; even the 304 landing pages with nominal capture record little activity (median observation span 0\,s). JavaScript activity was nevertheless recorded on 93\% of the affected pages, and 6.4\% of them carried page-denial signals, comparable to the 5--8\% of the other crawlers. Launching the same extension-equipped browser through Playwright reduced missing landing-page capture to 2\%, whereas the normally launched variant without simulated interaction missed 97\%. The participants, who used the same extension in an already running browser, lost capture on 24\% of their 374 landing pages (368 assigned plus 6 unassigned sites) and on 7\% of their internal pages, so the extension is least reliable on the first page that a fresh browser or tab loads, and \gui, which restarts Chrome before every site, exposes every landing page to this weakness: the extension must start and attach to the debugging interface before initial navigation. Under the exclusion rule in \cref{sec:instr}, including the 304 landing pages with nominal capture would lower its tracker ratio to its same-URL crawl from 1.69 to 1.64 and its tracker difference to \chr from +2.42 to +0.84 organizations per page. We did not substitute the Playwright-launched variant for the main run, because the launch path itself changes denial and WebGL activity (\cref{sec:access,sec:fingerprinting}).

\noindent\textbf{Fingerprinting diagnostics.} The raw logs contain WebGL calls in every arm, including 57 \gui and 93 \dom visits, so the collector does not suppress them, and \gui's normally launched and Playwright-launched variants ran byte-identical instrumentation in the same extension. On the 875 pages (430 sites) that both same-day variants recorded with JavaScript activity, the Playwright launch raised WebGL renderer queries from 1.3\% to 44.8\% and strict fingerprinting from 21.0\% to 46.4\%. On the 236 of these pages where the Playwright-launched variant recorded fingerprinting-like API calls by scripts from \texttt{pagead2.googlesyndication.com} or \texttt{ep1.adtrafficquality.google} (which commonly perform anti-fraud and automated-traffic checks), the normally launched variant recorded JavaScript from the same hosts on 216 pages and canvas calls on 199 pages, but no WebGL call. In the stealth bundle, Patchright alone left 3.7\% of pages with uncertain sequences and the BrowserForge profile alone 9.3\% (\chr: 8.3\%), and \ffs, which applies the overrides through Playwright, left 4.4\% (\ff: 4.8\%); device-metadata reads were unchanged by Patchright alone (13.9\% versus 14.5\% of pages) but fell to 4.3\% in \chrs, whereas WebGL renderer queries already fell with Patchright alone (14.3\% versus 22.4\%). \ff (\cref{sec:instr}) records less strict fingerprinting than \chr ($-$8.7\pp\ci{$-$10.2}{$-$7.3}, or $-$2.2\pp without WebGL), and it never triggers the device-metadata vector, which relies on APIs that Firefox does not implement.

\noindent\textbf{Trackers and consent platforms.} Tracker Radar~\cite{trackerradar} categories map to advertising (Advertising, Ad Motivated Tracking, Action Pixels), analytics (Analytics, Audience Measurement, Third-Party Analytics Marketing, Tag Manager), session replay, and social (Social Network, Social -- Share, Social -- Comment); a domain without a listed owner counts as its own organization. We treat OneTrust, Usercentrics, SFBX, CookieYes, Civic Computing, IAB Europe, Sirdata, ConsentManager, Digital Data Solutions, Cookie Information, Ringier, Sourcepoint, TrustArc, Didomi, Cybot, Osano, Termly, and iubenda as consent-management platforms.

\noindent\textbf{Fingerprinting.} Calls are grouped per visit by full script URL, so calls without a usable script URL cannot form sequences. The canvas vector requires text drawing followed within 500\,ms by an extraction other than JPEG or WebP, with \texttt{getImageData} covering at least $16\times16$ pixels; the font vector requires more than 20 font-family settings and more than 20 text measurements, or more than 20 families passed to \texttt{FontFaceSet.check}; and the WebGL vector requires \texttt{getParameter(37445)} or \texttt{getParameter(37446)}. The weak vectors are GPU-surface queries, a request for the debug-renderer extension without the unmasked query, local-font access, and device metadata (at least two of client hints, media-device enumeration, voices, storage estimate, \texttt{Intl} options, and hardware concurrency with device memory). A sequence is \emph{uncertain} when an extraction lacks a recorded draw or timestamp, or when a \texttt{getImageData} or \texttt{getParameter} call lacks its arguments. The legacy definition implements Englehardt and Narayanan's canvas, canvas-font, WebRTC, and audio rules~\cite{englehardt2016online}.

\noindent\textbf{Inference.} Whole-site bootstrap samples retain all matched pages. Sign-flip tests swap arms consistently within sites, assuming site-level exchangeability rather than guaranteed randomization. Weighting sites equally instead of pages changes none of the headline contrast signs (\cref{sec:stats}); escalation already uses equal site weights (\cref{sec:metrics}).

\section{Additional Results}
\label{app:extra}

\Cref{tab:contrasts,tab:accessmerged,tab:drift,tab:engextra,tab:functions} give planned contrasts, intervention-depth checks, collection drift, dwell checks, and additional trackers' ad-tech functions, respectively.

\noindent\textbf{Collection drift.} Same-URL crawls one week apart have a tracker ratio of 0.95 on shared internal pages; \chr variants 9--14 days after their reference yield 0.98--1.00, and \gui variants yield 0.93--0.97 relative to its unmodified reference 13 days later. Across three months, repeated \gui on the 450-site subset yields 1.01\ci{0.96}{1.07} on matched pages with unchanged internal denial. Landing denial changes by at most 1.6\pp across two weeks but rises 3.8\pp\ci{1.8}{5.8} across three months (\cref{tab:drift}). These checks support the comparison choices in \cref{sec:stats}.

\begin{table}[tbp]
\centering
\scriptsize
\setlength{\tabcolsep}{2.1pt}
\caption{Interventions by configuration. \emph{Landing}: signals on the 1,902 shared landing pages (\%; vis./pass.: visible/passive challenges). \emph{Escalation}: later-page minus landing-page rate (pp) on sites with ${\geq}K$ pages; $n$: sites at $K{=}6$. Bold: interval excludes zero.}
\label{tab:accessmerged}
\begin{tabular}{@{}lrrrrrrrrr@{}}
\toprule
& \multicolumn{4}{c}{Landing (\%)} & \multicolumn{3}{c}{Escalation, denial} & Hard & \\
\cmidrule(lr){2-5}\cmidrule(lr){6-8}\cmidrule(lr){9-9}
Configuration & Den. & Hard & Vis. & Pass. & $K{=}5$ & 6 & 8 & 6 & $n$ \\
\midrule
\input{tables/tab_access_merged_rows.tex} 
\bottomrule
\end{tabular}
\end{table}

\begin{table}[tbp]
\centering
\footnotesize
\setlength{\tabcolsep}{2.4pt}
\caption{Drift between repeated collections: change in page denial (pp, variant minus reference) and tracker ratio on matched pages. Internal-page changes for Patchright and the Playwright launch are configuration effects. Bold: interval excludes zero (one for ratios).}
\label{tab:drift}
\begin{tabular}{@{}lrrrr@{}}
\toprule
Repeat & Days & Landing & Internal & Trackers \\
\midrule
\input{tables/tab_drift_rows.tex} 
\bottomrule
\end{tabular}
\end{table}

\begin{table}[tbp]
\centering
\footnotesize
\setlength{\tabcolsep}{2.5pt}
\caption{Dwell robustness: tracker-organization ratio, agent over its same-URL crawl (95\% intervals). \emph{Both active at $T$}: pairs still recording at $T$, counting organizations first seen by $T$.}
\label{tab:engextra}
\begin{tabular}{@{}lrrrr@{}}
\toprule
& \multicolumn{2}{c}{\dom} & \multicolumn{2}{c}{\gui} \\
\cmidrule(lr){2-3}\cmidrule(lr){4-5}
Population & Ratio & Pages & Ratio & Pages \\
\midrule
\input{tables/tab_engagement_extra_rows.tex} 
\bottomrule
\end{tabular}
\end{table}
\begin{table}[tbp]
\centering
\footnotesize
\setlength{\tabcolsep}{3pt}
\caption{The agents' additional trackers by ad-tech function: organizations recorded only by the agent minus only by its same-URL crawl, per matched page, and share of the total (\%). Functions assigned manually for the 45 largest contributors.}
\label{tab:functions}
\begin{tabular}{@{}lrrrr@{}}
\toprule
& \multicolumn{2}{c}{\dom} & \multicolumn{2}{c}{\gui} \\
\cmidrule(lr){2-3}\cmidrule(lr){4-5}
Function & Per page & \% & Per page & \% \\
\midrule
\input{tables/tab_functions_rows.tex} 
\bottomrule
\end{tabular}
\end{table}

%% file: tables/tab_contrasts_rows.tex
\multicolumn{8}{@{}l}{\emph{Stealth vs.\ base configuration}} \\
Chromium-S $-$ Chromium & 7,109 & \textbf{$-$4.2} {\scriptsize [$-$5.5, $-$3.0]} & \textbf{$-$5.1} {\scriptsize [$-$6.5, $-$3.8]} & 6,757 & \textbf{$-$0.27} {\scriptsize [$-$0.34, $-$0.19]} & \textbf{$-$10.9} {\scriptsize [$-$12.6, $-$9.4]} & \textbf{$-$4.8} {\scriptsize [$-$6.0, $-$3.7]} \\
Firefox-S $-$ Firefox & 7,247 & +0.4 {\scriptsize [0.0, 0.9]} & +0.3 {\scriptsize [$-$0.2, 0.8]} & 6,946 & +0.03 {\scriptsize [$-$0.02, 0.08]} & $-$0.5 {\scriptsize [$-$1.1, 0.2]} & $-$0.4 {\scriptsize [$-$1.0, 0.1]} \\
\midrule
\multicolumn{8}{@{}l}{\emph{Browser engine}} \\
Firefox $-$ Chromium & 7,121 & +0.4 {\scriptsize [$-$0.2, 0.9]} & +0.1 {\scriptsize [$-$0.5, 0.8]} & 6,821 & \textbf{$-$0.29} {\scriptsize [$-$0.34, $-$0.23]} & \textbf{$-$8.7} {\scriptsize [$-$10.2, $-$7.3]} & \textbf{$-$2.2} {\scriptsize [$-$3.0, $-$1.3]} \\
\midrule
\multicolumn{8}{@{}l}{\emph{Agentic vs.\ non-agentic crawlers}} \\
browser-use $-$ Chromium & 3,464 & \textbf{$-$3.5} {\scriptsize [$-$4.7, $-$2.4]} & \textbf{$-$4.8} {\scriptsize [$-$6.1, $-$3.5]} & 3,141 & \textbf{+2.05} {\scriptsize [1.78, 2.33]} & \textbf{$-$7.3} {\scriptsize [$-$9.3, $-$5.3]} & \textbf{+3.1} {\scriptsize [1.7, 4.6]} \\
SuperScraper $-$ Chromium & 2,565 & \textbf{$-$1.9} {\scriptsize [$-$2.8, $-$1.0]} & \textbf{$-$4.0} {\scriptsize [$-$5.2, $-$2.7]} & 572 & \textbf{+2.42} {\scriptsize [1.83, 3.04]} & \textbf{$-$8.1} {\scriptsize [$-$10.1, $-$6.0]} & \textbf{+3.2} {\scriptsize [1.6, 4.8]} \\
browser-use $-$ Chromium-S & 3,376 & $-$0.9 {\scriptsize [$-$1.7, $-$0.3]} & $-$0.4 {\scriptsize [$-$1.2, 0.3]} & 3,008 & \textbf{+2.87} {\scriptsize [2.58, 3.18]} & \textbf{+5.7} {\scriptsize [4.1, 7.4]} & \textbf{+9.6} {\scriptsize [8.3, 11.0]} \\
SuperScraper $-$ Chromium-S & 2,526 & $-$1.1 {\scriptsize [$-$2.0, $-$0.3]} & $-$0.7 {\scriptsize [$-$1.7, 0.3]} & 525 & \textbf{+2.16} {\scriptsize [1.62, 2.75]} & \textbf{+6.1} {\scriptsize [4.4, 8.0]} & \textbf{+9.8} {\scriptsize [8.2, 11.4]} \\
\midrule
\multicolumn{8}{@{}l}{\emph{The two agentic crawlers}} \\
browser-use $-$ SuperScraper & 2,859 & 0.0 {\scriptsize [$-$0.6, 0.5]} & $-$0.2 {\scriptsize [$-$1.1, 0.5]} & 883 & $-$0.09 {\scriptsize [$-$0.38, 0.19]} & +0.3 {\scriptsize [$-$1.2, 1.9]} & +0.3 {\scriptsize [$-$1.2, 1.9]} \\
\midrule
\multicolumn{8}{@{}l}{\emph{Agentic crawlers vs.\ their same-URL crawls}} \\
browser-use $-$ same-URL crawl & 12,089 & $-$0.6 {\scriptsize [$-$1.1, 0.0]} & \textbf{$-$2.8} {\scriptsize [$-$3.8, $-$1.8]} & 11,435 & \textbf{+1.99} {\scriptsize [1.72, 2.26]} & \textbf{$-$3.0} {\scriptsize [$-$4.8, $-$1.2]} & +1.0 {\scriptsize [$-$0.6, 2.6]} \\
SuperScraper $-$ same-URL crawl & 5,815 & \textbf{$-$2.6} {\scriptsize [$-$3.5, $-$1.8]} & \textbf{$-$4.0} {\scriptsize [$-$5.3, $-$2.6]} & 3,759 & \textbf{+2.95} {\scriptsize [2.53, 3.36]} & $-$1.7 {\scriptsize [$-$3.8, 0.4]} & \textbf{+2.8} {\scriptsize [1.0, 4.6]} \\
\midrule
\multicolumn{8}{@{}l}{\emph{Residential proxy vs.\ none}} \\
browser-use: proxy $-$ none & 5,108 & \textbf{+0.7} {\scriptsize [0.3, 1.2]} & \textbf{+1.0} {\scriptsize [0.5, 1.5]} & 4,734 & +0.01 {\scriptsize [$-$0.17, 0.18]} & $-$0.6 {\scriptsize [$-$1.6, 0.3]} & $-$0.6 {\scriptsize [$-$1.6, 0.3]} \\
SuperScraper: proxy $-$ none & 2,631 & \textbf{+2.2} {\scriptsize [1.4, 2.9]} & \textbf{+2.4} {\scriptsize [1.5, 3.3]} & 635 & $-$0.41 {\scriptsize [$-$0.75, $-$0.07]} & +1.0 {\scriptsize [$-$0.3, 2.2]} & +1.0 {\scriptsize [$-$0.3, 2.2]} \\

%% file: figures/fig_superscraper.tex
\begin{tikzpicture}[
  font=\sffamily\scriptsize, text=black!65,
  >={Stealth[length=1.5mm, width=1.6mm]},
  badge/.style={circle, fill=#1, minimum size=6.2mm, inner sep=0pt, text=white, font=\normalsize},
  loop/.style={->, line width=1pt, draw=cSS!45},
  lbl/.style={inner sep=0pt, align=#1},
  card/.style={fill=black!2.5, draw=black!22, rounded corners=3pt, line width=0.5pt},
  link/.style={->, line width=0.6pt, draw=black!40},
  aside/.style={line width=0.6pt, draw=black!35, densely dotted},
]
\definecolor{cSS}{HTML}{4A3AA7}
\definecolor{cEnv}{HTML}{5F6670}
\def\R{1.6}
\def\ttl#1#2{{\footnotesize\bfseries{\color{cSS!55}#1}\enspace{\color{black!85}#2}}}
\def\sscard#1#2#3#4#5{%
  \draw[card] (#1,#2) rectangle ++(2.98,-1.2);
  \node[anchor=base west, inner sep=0pt, text=black!85] at (#1+0.13,#2-0.3)
    {{\color{cSS}\faIcon{#3}}\enspace\textbf{#4}};
  \foreach \ic/\tx [count=\k] in {#5} {
    \node[anchor=base, inner sep=0pt, text=black!40, font=\tiny] at (#1+0.24,#2-0.3-0.25*\k) {\faIcon{\ic}};
    \node[anchor=base west, inner sep=0pt] at (#1+0.43,#2-0.3-0.25*\k) {\tx};
  }%
}

\fill[cSS!4] (0,0) circle (\R);
\node[badge=cSS] (obs) at (210:\R) {\faIcon{camera}};
\node[badge=cSS] (prs) at (150:\R) {\faIcon{vector-square}};
\node[badge=cSS] (dec) at ( 90:\R) {\faIcon{brain}};
\node[badge=cSS] (val) at ( 30:\R) {\faIcon{clipboard-check}};
\node[badge=cSS] (act) at (-30:\R) {\faIcon{mouse-pointer}};
\node[badge=cEnv] (chr) at (-90:\R) {\faIcon{chrome}};
\foreach \a in {210,150,90,30,-30,-90}
  \draw[loop] (\a-12.5:\R) arc[start angle=\a-12.5, end angle=\a-47.5, radius=\R];

\node[lbl=right, anchor=east] at ([xshift=-1.3mm]obs.west) {\ttl{1}{Observe}\\screenshot + URL};
\node[lbl=right, anchor=east] at ([xshift=-1.3mm]prs.west) {\ttl{2}{Parse}\\OmniParser\\labeled boxes};
\node[lbl=center, anchor=south] at ([yshift=1.2mm]dec.north) {\ttl{3}{Decide}\\Qwen3-VL-8B\\GBNF grammar};
\node[lbl=left, anchor=west] at ([xshift=1.3mm]val.east) {\ttl{4}{Check}\\well-formed JSON\\argument checks};
\node[lbl=left, anchor=west] at ([xshift=1.3mm]act.east) {\ttl{5}{Act}\\OS mouse + keyboard};
\node[lbl=center, anchor=north] at ([yshift=-1.2mm]chr.south) {{\footnotesize\bfseries\color{black!85}Chrome}\\ordinary profile};

\node[fill=white, draw=cSS!35, rounded corners=3pt, line width=0.5pt, align=center, inner sep=2.5pt]
  (st) at (0,-0.1) {{\color{cSS}\small\faIcon{history}}\\[0.5pt]{\bfseries\color{black!85}Session state}\\visited URLs\\last 10 actions};
\draw[->, line width=0.7pt, draw=cSS!45, densely dashed] (st) -- (dec);

\sscard{-4.18}{2.6}{bullseye}{Task}{list-ol/10 distinct URLs, window-close/dismiss overlays, link/{internal links, no ads}}
\sscard{1.2}{2.6}{flag-checkered}{Stop when}{check/10 URLs reached, redo/{5 steps, same URL}, hourglass-end/60 iterations}
\sscard{-4.18}{-1.4}{puzzle-piece}{Extension}{code/JS instrumentation, exchange-alt/response capture, eye-slash/not seen by agent}
\sscard{1.2}{-1.4}{random}{Behavior simulation}{route/cursor trajectories, stopwatch/randomized timing, arrows-alt-v/scroll after URL change}

\draw[link] (-1.2,2.0) -- (dec.160);
\draw[link, densely dashed] (chr.195) -- (-1.2,-1.85);
\draw[aside] (act.south) -- (act.south |- 0,-1.4);
\end{tikzpicture}

%% file: tables/tab_access_merged_rows.tex
Chromium & 6.0 & 11.0 & 0.3 & 8.6 & \textbf{+8.9} & \textbf{+10.1} & \textbf{+12.6} & \textbf{+4.6} & 879 \\
Chromium-S & 6.5 & 8.3 & 0.5 & 2.2 & \textbf{+1.2} & \textbf{+1.2} & \textbf{+1.6} & +0.3 & 868 \\
Firefox & 7.1 & 10.9 & 0.1 & 6.5 & \textbf{+5.2} & \textbf{+4.9} & \textbf{+6.3} & \textbf{+1.8} & 701 \\
Firefox-S & 8.1 & 11.8 & 0.1 & 3.6 & \textbf{+5.2} & \textbf{+4.4} & \textbf{+5.9} & \textbf{+2.5} & 697 \\
browser-use & 5.3 & 7.5 & 0.5 & 17.5 & $-$0.4 & $-$0.4 & $-$0.2 & $-$0.6 & 842 \\
browser-use (proxy) & 6.5 & 9.4 & 0.5 & 14.2 & $-$0.2 & $-$0.3 & $-$0.9 & \textbf{$-$1.4} & 844 \\
SuperScraper & 5.5 & 7.9 & 0.5 & 18.1 & \textbf{$-$2.1} & \textbf{$-$1.7} & \textbf{$-$1.7} & $-$1.5 & 344 \\
SuperScraper (proxy) & 8.1 & 11.0 & 0.4 & 20.3 & \textbf{$-$2.7} & \textbf{$-$2.6} & \textbf{$-$2.9} & \textbf{$-$3.4} & 364 \\

%% file: tables/tab_drift_rows.tex
Same-URL crawls (internal pages) & 7 & -- & -- & \textbf{0.95} \\
Chromium $\to$ Patchright & 11 & $-$0.3 & \textbf{$-$6.4} & 0.99 \\
Chromium + BrowserForge & 14 & $-$0.2 & +0.3 & \textbf{0.98} \\
Chromium + click navigation & 9 & +0.1 & +0.5 & 1.00 \\
SuperScraper $-$ behavior sim. & 13 & +0.2 & +0.0 & 0.97 \\
SuperScraper + Playwright launch & 13 & $-$1.6 & +1.3 & 0.93 \\
SuperScraper, repeated & 91 & \textbf{+3.8} & +0.0 & 1.01 \\

%% file: tables/tab_engagement_extra_rows.tex
Both active at 2\,s, first 2\,s & 2.23\,{\scriptsize[2.09, 2.38]} & 9,999 & 2.65\,{\scriptsize[2.45, 2.85]} & 3,266 \\
Both active at 5\,s, first 5\,s & 2.43\,{\scriptsize[2.26, 2.62]} & 8,801 & 2.55\,{\scriptsize[2.39, 2.76]} & 2,397 \\
Both active at 10\,s, first 10\,s & 1.64\,{\scriptsize[1.53, 1.76]} & 3,922 & 1.89\,{\scriptsize[1.76, 2.04]} & 1,303 \\
Both active at 20\,s, first 20\,s & 1.04\,{\scriptsize[0.98, 1.11]} & 640 & 1.43\,{\scriptsize[1.34, 1.53]} & 197 \\
Agent span $\leq$ same-URL crawl span & 1.04\,{\scriptsize[0.95, 1.14]} & 1,932 & 1.10\,{\scriptsize[1.03, 1.17]} & 481 \\
Internal pages only & 1.48\,{\scriptsize[1.40, 1.58]} & 9,804 & 1.69\,{\scriptsize[1.57, 1.81]} & 3,759 \\

%% file: tables/tab_functions_rows.tex
Supply-side platforms and exchanges & +0.60 & 30 & +0.90 & 30 \\
Platform ads, pixels and analytics & +0.43 & 22 & +0.59 & 20 \\
Ad verification and measurement & +0.21 & 10 & +0.42 & 14 \\
Demand-side and retargeting & +0.16 & 8 & +0.19 & 6 \\
Content recommendation & +0.07 & 4 & +0.15 & 5 \\
Identity and audience data & $-$0.02 & $-$1 & +0.02 & 1 \\
All other organizations & +0.55 & 27 & +0.69 & 23 \\